\documentclass[preprint,12pt]{elsarticle}

\usepackage{bm}% bold math

\biboptions{sort&compress}
\usepackage{amsmath,bbold,amssymb,epsfig,bm,mathrsfs}
\usepackage{feynmp,color,slashed,nicefrac,exscale,multirow,graphicx}
\usepackage{times,txfonts}
\usepackage{epstopdf}

\usepackage[dvipdfm,bookmarks=true,colorlinks,%
            citecolor=blue,linkcolor=blue,hypertex, %
            breaklinks=true]{hyperref}

\journal{Physics Letters B}

\begin{document}

\hyphenpenalty=6000
\tolerance=1000

\begin{frontmatter}



\title{Strength of pairing interaction for hyperons in multistrangeness hypernuclei}


\author[ITP,UCAS]         {Yu-Ting Rong}
\author[PKU]              {Pengwei Zhao}
\author[ITP,UCAS,CTNP,HNU]{Shan-Gui Zhou\corref{cor1}}
\ead{sgzhou@itp.ac.cn}
\cortext[cor1]{Corresponding author.}

\address[ITP] {CAS Key Laboratory of Theoretical Physics,
	           Institute of Theoretical Physics, Chinese Academy of Sciences, 
               Beijing 100190, China}
\address[UCAS]{School of Physical Sciences, 
               University of Chinese Academy of Sciences, Beijing 100049, China}
\address[PKU] {State Key Laboratory of Nuclear Physics and Technology, 
               School of Physics, Peking University, Beijing 100871, China}
\address[CTNP]{Center of Theoretical Nuclear Physics, National Laboratory of 
               Heavy Ion Accelerator, Lanzhou, 730000, China}
\address[HNU] {Synergetic Innovation Center for Quantum Effects and Application, 
               Hunan Normal University, Changsha, 410081, China}

\begin{abstract}
Pairing correlations play a very important role in atomic nuclei. 
Although several effective pairing interactions have been used in mean field 
calculations for nucleons, little is known about effective pairing interactions 
for hyperons. 
Based on the quark model, we propose a relationship between effective pairing 
interactions for hyperons and for nucleons; 
e.g., for $\Lambda$s, the strength of the pairing interaction is $4/9$ of that 
for nucleons. 
A separable pairing force of finite range which has been widely applied to
describing pairing correlations in normal nuclei is used to investigate pairing 
effects in multi-$\Lambda$ Ca, Sn and Pb hypernuclei.
\end{abstract}


\end{frontmatter}

%\section{Introduction} 

Since the discovery of the first $\Lambda$ hypernucleus from cosmic rays \cite{Danysz1953_PhilosMag44-348}, 
the study of hypernuclei has been one of very interesting topics in nuclear physics
\cite{Hashimoto2006_PPNP57-564,Hiyama2010_PTPSuppl185-106,Hiyama2010_PTPSuppl185-152,Tamura2012_PTEP2012-02B012,Feliciello2015_RPP78-096301,Gal2016_RMP88-035004,
Chen2019_JPG46-125106,Nogga2019_AIPCP2130-030004}. 
Most of the observed hypernuclei are of single-$\Lambda$. 
So far there is only one confirmed double-$\Lambda$ hypernucleus 
$_{\Lambda\Lambda}^{~~6}$He \cite{Takahashi2001_PRL87-212502} and two candidates $^{~13}_{\Lambda\Lambda}$B \cite{Aoki2009_NPA828-191} and 
$^{~10}_{\Lambda\Lambda}$Be \cite{Ahn2013_PRC88-014003}. 
Nevertheless, many theoretical efforts have been devoted to investigating the 
structure of more double-$\Lambda$ and even multistrangeness ($-S\ge 3$) hypernuclei 
\cite{Miyahara1983_PTP69-1717,
Lu2002_CPL19-1775,
Lu2008_CPL25-3613,Shoeb2009_JPG36-045104,
Gal2011_PLB701-342,
Margueron2017_PRC96-054317,
Gueven2018_PRC98-014318,
Tanimura2019_PRC99-034324}.

Hypernuclei are unique quantum many-body systems for the investigation of 
hyperon-nucleon ($YN$) and hyperon-hyperon ($YY$) interactions which are, 
in turn, crucial for understanding the hypernuclear structure as well as 
the hypernuclear matter and properties of neutron stars. 
With the strangeness degree of freedom, a hyperon can move deep inside the 
nucleus and serve as an impurity for probing nuclear properties that are not 
accessible by conventional methods developed for normal nuclei. 
Lots of many-body techniques for normal nuclei,
including 
various mean-field models \cite{%
Lu2003_EPJA17-19,
Zhou2005_PRL95-051101,
Shen2006_PTP115-325,
Zhou2009_SciChinaG52-1548,
Schulze2010_PTP123-569,
Win2011_PRC83-014301,
Lu2011_PRC84-014328,
Song2011_CPL28-092101,
Xu2012_JPG39-085107,
Lu2014_PRC89-044307,
Mei2015_PRC91-064305,
Sun2016_PRC94-064319,
Cui2017_PRC95-024323,
Margueron2017_PRC96-054317,
Wu2017_PRC95-034309,
Sun2017_PRC96-044312,
Li2018_PRC97-034302,
Mei2018_PRC97-064318,
Gueven2018_PRC98-014318,
Liu2018_PRC98-024316,
Tanimura2019_PRC99-034324,
Xia2019_SciChinaPMA62-042011}, %, etc.
have been extended to hypenuclei. 
In particular, the relativistic mean-field (RMF) models 
\cite{Serot1986_ANP16-1,
Reinhard1989_RPP52-439,Ring1996_PPNP37-193,Bender2003_RMP75-121,
Vretenar2005_PR409-101,Meng2006_PPNP57-470,Niksic2011_PPNP66-519,
Liang2015_PR570-1,Meng2015_JPG42-093101,
Zhou2016_PS91-063008}
which have been very successful in describing normal nuclei in the whole nuclear 
chart are also used extensively to study hypernuclei. 

In the RMF models, one needs effective interactions both for the particle-hole 
($ph$) and particle-particle ($pp$) channels.
For normal nuclei, a large amount of effective interactions for the $ph$ channel 
have been proposed, see, e.g., Refs.~\cite{Nikolaus1992_PRC46-1757,
Sharma1993_PLB312-377,Sugahara1994_NPA579-557,Lalazissis1997_PRC55-540,
Long2004_PRC69-034319,Lalazissis2005_PRC71-024312,Niksic2008_PRC78-034318,
Lalazissis2009_PLB671-36,Zhao2010_PRC82-054319}. 
Meanwhile, both zero-range and finite-range pairing forces have been used 
in the $pp$ channel \cite{Meng1998_NPA635-3,Tian2009_PLB676-44,Robledo2018_JPG46-013001}.
For hypernuclei, $YN$ and $YY$ interactions in the $ph$ channel can be either
obtained by fitting experimental data or estimated with the naive quark model  \cite{Sugahara1994_PTP92-803,Mares1994_PRC49-2472,Ma1996_NPA608-305,
Wang2013_CTP60-479,Wang2010_PRC81-025801,Tanimura2012_PRC85-014306,
Xu2012_JPG39-085107}. 
However, effective interactions for hyperons in the $pp$ channel are much less known. 
In this Letter, we propose a way to estimate the strength of pairing interactions 
for hyperons based on the meson exchange picture and the naive quark model. 

From the quark model we know that nucleons consist of three u/d quarks and 
hyperons consist of, besides u/d quarks, one or more s quarks. 
Next we use $n_{\mathrm{u/d}}$ to label the number of u/d quarks in a baryon 
(a nucleon or a hyperon) and define $g_{BM}$ to be the coupling constant of 
a non-strangeness meson $M$ ($\sigma$, $\omega$, $\rho$, $\cdots$) to a baryon 
$B$ ($N$, $\Lambda$, $\cdots$). 
According to the OZI rule 
only u/d quarks are involved in the coupling of a non-strangeness meson 
to a baryon at the tree level. 
Therefore the following relation holds: $g_{YM} = n_{\mathrm{u/d}}/3 \cdot g_{NM}$. 
Similar discussions have been made with the quark-meson coupling model 
\cite{Tsushima1997_PLB411-9}. 
If $g_{NM}$ is known, one can readily get $g_{YM}$. 
For example, since $n_{\mathrm{u/d}} = 2$ in $\Lambda$s,  
$g_{\Lambda M} = 2/3 \cdot g_{NM}$; 
this has been proposed and used in the study of $\Lambda$-hypernuclei 
\cite{Pirner1979_PLB85-190,Dover1984_PPNP12-171,Mares1994_PRC49-2472,
Sugahara1994_PTP92-803,Schaffner1994_AP235-35}.

The exchange of the meson $M$ between two baryons $B_1$ and $B_2$ results 
in an interaction with the strength proportional to $g_{B_1M} g_{B_2M}$ \cite{Yukawa1935_ProcPhysMathSocJapan17-48}. 
For single-$\Lambda$ hypernuclei, the central potential for the $\Lambda$
in the mean field generated by nucleons is proportional to $g_{NM} g_{\Lambda M}$ 
while that for nucleons is proportional to $g^2_{NM}$, leading to 
the well known observation that the depth of the potential for the $\Lambda$ 
is roughly 2/3 of that for nucleons \cite{Bouyssy1976_PLB64-276,
Dover1987_ProcISMEP257,Millener1988_PRC38-2700,Rodriguez-Sanchez2018_PRC98-021602R}.

The strength for the $YY$ interaction is proportional to $g^2_{YM}$, i.e., 
$n^2_{\mathrm{u/d}}/9$ $\left( = g^2_{YM}/g^2_{NM} \right)$ of that for the $NN$ interaction. 
Since the pairing force is the residual of the two-body $BB$ interaction, 
the ratio $n^2_{\mathrm{u/d}}/9$ holds also between the strength of the pairing 
interaction for hyperons and that for nucleons. 
Note that mesons consisting of strange quarks may be exchanged between hyperons
and result in possible deviations of this ratio from $n^2_{\mathrm{u/d}}/9$. 
In this work we restrict our discussions in the framework of conventional 
RMF models with non-strangeness mesons. 

As an illustration, we use the relativistic Hartree-Bogoliubov (RHB) theory 
to study the effects of the $\Lambda\Lambda$ pairing in multistrangeness hypernuclei. 
The RHB theory provides a unified description of the relativistic mean field 
and the pairing correlations via the Bogoliubov transformation. 
The RHB equation consists of the single particle Hamiltonian $h_B$ and 
the pairing field $\Delta$ \cite{Ring1996_PPNP37-193,Kucharek1991_ZPA339-23}
\begin{equation}
\label{eq:rhb}
 \int d^{3}\bm{r}^{\prime}
  \left( \begin{array}{cc} h_B-\lambda  &  \Delta                      \\ 
                          -\Delta^{*}   & -h_B+\lambda \end{array} 
  \right)
  \left( \begin{array}{c} U_{k} \\ V_{k} \end{array} \right)
 = E_{k}
  \left( \begin{array}{c} U_{k} \\ V_{k} \end{array} \right),
\end{equation}
where $\lambda$ is the Fermi energy and $E_k$ and $(U_k,V_k)^T$ are the quasi-particle 
energy and wave function, respectively. 
%\begin{equation}
% \begin{pmatrix}  h_B - \lambda &  \Delta        \\ 
%                 -\Delta^*      & -h_B^*+\lambda 
% \end{pmatrix}
% \begin{pmatrix} \psi_U \\ \psi_V 
% \end{pmatrix}
% =
% E 
% \begin{pmatrix} \psi_U \\ \psi_V 
% \end{pmatrix},
%\end{equation}

For the $ph$ channel, the Dirac Hamiltonian for nucleons has been established 
\cite{Serot1986_ANP16-1,Reinhard1989_RPP52-439,Ring1996_PPNP37-193,
Bender2003_RMP75-121,Vretenar2005_PR409-101,Meng2006_PPNP57-470,
Niksic2011_PPNP66-519,Liang2015_PR570-1,Meng2015_JPG42-093101,
Zhou2016_PS91-063008} 
and that for $\Lambda$s reads \cite{Schaffner1994_AP235-35}
\begin{equation}
 h_\Lambda(\bm{r}) = \bm{\alpha} \cdot \bm{p} + V_\Lambda(\bm{r}) + 
                     \beta \left( m_\Lambda + S_\Lambda(\bm{r}) \right) + 
                     T_\Lambda(\bm{r}),
\end{equation}
where the scalar, vector and tensor potentials are
\begin{equation}
\begin{split}
 S_\Lambda(\bm{r}) &= g_{\sigma\Lambda} \sigma  (\bm{r}), \\
 V_\Lambda(\bm{r}) &= g_{\omega\Lambda} \omega_0(\bm{r}), \\
 T_\Lambda(\bm{r}) &=-\frac{f_{\omega\Lambda\Lambda}}{2m_\Lambda} 
                      \beta (\bm{\alpha} \cdot \bm{p}) \omega_0(\bm{r}).
\end{split}
\end{equation}
The tensor potential $T_\Lambda (\bm{r})$ is included to achieve the small 
spin-orbit splitting for the $\Lambda$ \cite{Brueckner1978_PLB79-157,
Noble1980_PLB89-325,Jennings1990_PLB246-325}. 
% ,Ajimura2001_PRL86-4255,Kohri2002_PRC65-034607}.
We adopt two effective interactions 
NLSH-A \cite{Sharma1993_PLB312-377,Mares1994_PRC49-2472} and
PK1-Y1 \cite{Long2004_PRC69-034319,Wang2013_CTP60-479} which have been
extensively used in the study of $\Lambda$ hypernuclei.

In the $pp$ channel, the pairing potential reads
\begin{equation}
\begin{aligned}
 \Delta(\bm{r}_{1}\sigma_{1},\bm{r}_{2}\sigma_{2})
 = & \int d^{3}\bm{r}_{1}^{\prime} d^{3}\bm{r}_{2}^{\prime} 
     \sum_{\sigma_{1}^{\prime}\sigma_{2}^{\prime}}
     V(\bm{r}_{1}         \sigma_{1},          \bm{r}_{2}         \sigma_{2},
       \bm{r}_{1}^{\prime}\sigma_{1}^{\prime}, \bm{r}_{2}^{\prime}\sigma_{2}^{\prime}) 
     \kappa 
      (\bm{r}_{1}^{\prime}\sigma_{1}^{\prime}, 
       \bm{r}_{2}^{\prime}\sigma_{2}^{\prime}),
\end{aligned}
\end{equation}
where 
$V$ is the effective pairing interaction and $\kappa$ is the pairing tensor
\begin{equation}
 \kappa(\bm{r_1}\sigma_1,\bm{r_2}\sigma_2) 
 = \sum_{k>0} V^{*}_{k}(\bm{r_1}\sigma_1) U_{k}(\bm{r_2}\sigma_2).
\end{equation}
We use the separable pairing force of finite range 
proposed by Tian et al. \cite{Tian2006_CPL23-3226,Tian2009_PLB676-44,
Tian2009_PRC80-024313,Tian2009_PRC79-064301}
\begin{equation}
\label{eq:separable}
 V(\bm{r}_{1}         \sigma_{1},         \bm{r}_{2}         \sigma_{2},
   \bm{r}_{1}^{\prime}\sigma_{1}^{\prime},\bm{r}_{2}^{\prime}\sigma_{2}^{\prime})
 = 
 -G \delta(\bm{R}-\bm{R}^{\prime}) P(\bm{r}) P(\bm{r}^{\prime}) 
 \frac{1-P_{\sigma}}{2},
\end{equation}
where $G$ is the pairing strength and 
$\bm{R}=(\bm{r}_{1}+\bm{r}_{2}) / 2$ and $\bm{r}=\bm{r}_{1}-\bm{r}_{2}$
are the center of mass and relative coordinates, respectively. 
$P(\bm{r})$ denotes the Gaussian function,
\begin{equation}
 P(\bm{r}) = {\left( 4\pi a^2 \right)^{-3/2}} e^{-{r^2}/{4 a^2}},
\end{equation}
where $a$ is the effective range of the pairing force.
For nucleons, the pairing strength $G_N=728$ MeV$\cdot$fm$^3$ and 
the effective range $a=0.644$ fm have been obtained by fitting 
the momentum dependence of the pairing gap in the nuclear matter 
calculated from the Gogny force. 
According to our proposal discussed before, the pairing strength for 
$\Lambda$s is taken to be $G_\Lambda = 4/9 \cdot G_N$. 

\begin{figure}[!htbp]
\centering
\includegraphics[width=0.32\textwidth]{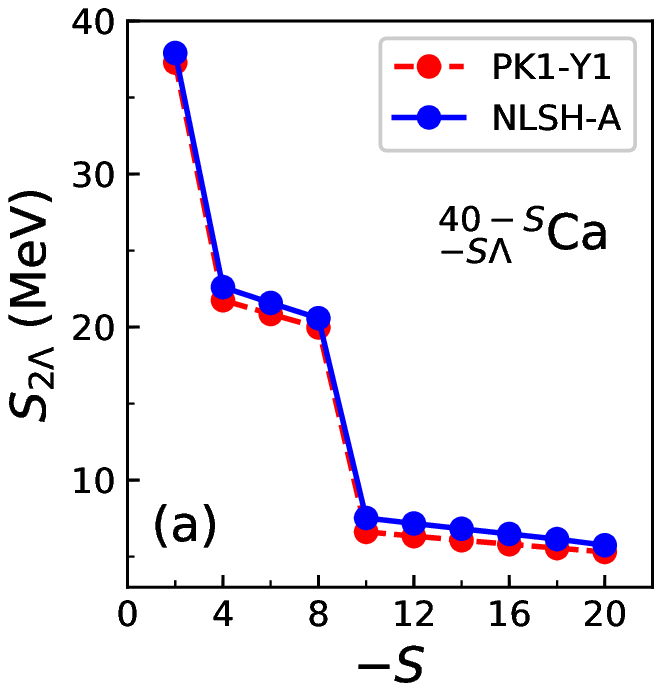}
\includegraphics[width=0.32\textwidth]{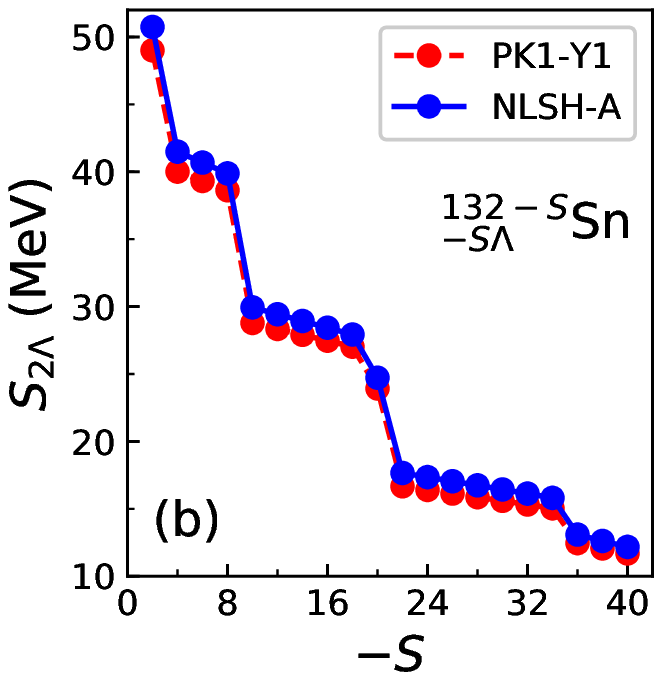}
\includegraphics[width=0.32\textwidth]{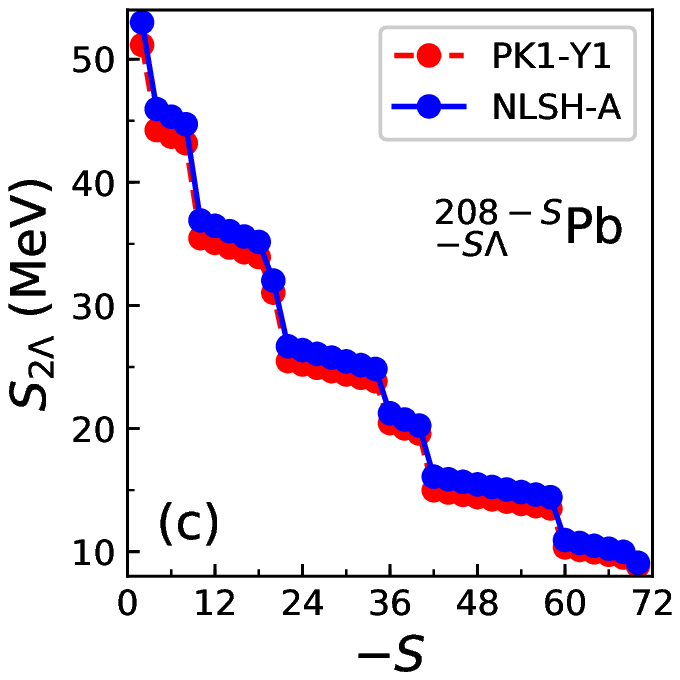}
\caption{(Color online) 
The two-Lambda separation energy as a function of the strangeness number $-S$ for
(a)  $^{40-S}_{-S\Lambda}$Ca ($-S=$ 0--20), 
(b) $^{132-S}_{-S\Lambda}$Sn ($-S=$ 0--40) and 
(c) $^{208-S}_{-S\Lambda}$Pb ($-S=$ 0--70) obtained in the MDC-RHB calculations. 
The effective interactions PK1-Y1 and NLSH-A are used for the $ph$ channel and 
the separable pairing force of finite range with the pairing strength 
$G_\Lambda= 4/9 \cdot G_N = 323.56$ MeV$\cdot$fm$^3$ is used for $pp$ channel.  
}\label{fig:2L_separation_energy}
\end{figure}

We have carried out calculations with the multidimensionally-constrained 
(MDC) RHB theory \cite{Zhao2017_PRC95-014320}, one of the recently developed
MDC covariant density functional theories (MDC-CDFTs) \cite{Zhou2016_PS91-063008,
Lu2012_PRC85-011301R,Lu2014_PRC89-014323,Zhao2017_PRC95-014320,
Meng2020_SciChinaPMA63-212011}. 
For simplicity, we choose doubly-magic $^{40}$Ca, $^{132}$Sn and $^{208}$Pb 
as the core nuclei and study even-even-even hypernuclei 
 $^{40-S}_{-S\Lambda}$Ca ($-S = 0$--20), 
$^{132-S}_{-S\Lambda}$Sn ($-S = 0$--40) and 
$^{208-S}_{-S\Lambda}$Pb ($-S = 0$--70). 
All these hypernuclei are spherical and have vanishing neutron and proton 
pairing gaps according to our MDC-RHB calculations.
In Fig.~\ref{fig:2L_separation_energy}, the two-Lambda separation energies 
$S_{2\Lambda}$ are shown for them. 
One can find that $S_{2\Lambda}$ decreases monotonically with the number of 
$\Lambda$s increasing.
As far as the two-Lambda separation energy is concerned, 
at least 20, 40 and 70 $\Lambda$s can be bound to the core nuclei $^{40}$Ca,
$^{132}$Sn and $^{208}$Pb, respectively.
There are sudden drops in $S_{2\Lambda}$ when $-S =$ 2, 8, 20, 34, 40 and 58.
These numbers are magic numbers for $\Lambda$s. 
Since the spin-orbit splitting is very small in the single $\Lambda$ spectrum, 
these numbers actually correspond to shell closures or sub-closures in the
single particle level scheme of a harmonic oscillator potential.

\begin{figure}[!htbp]
\centering
\includegraphics[width=0.32\textwidth]{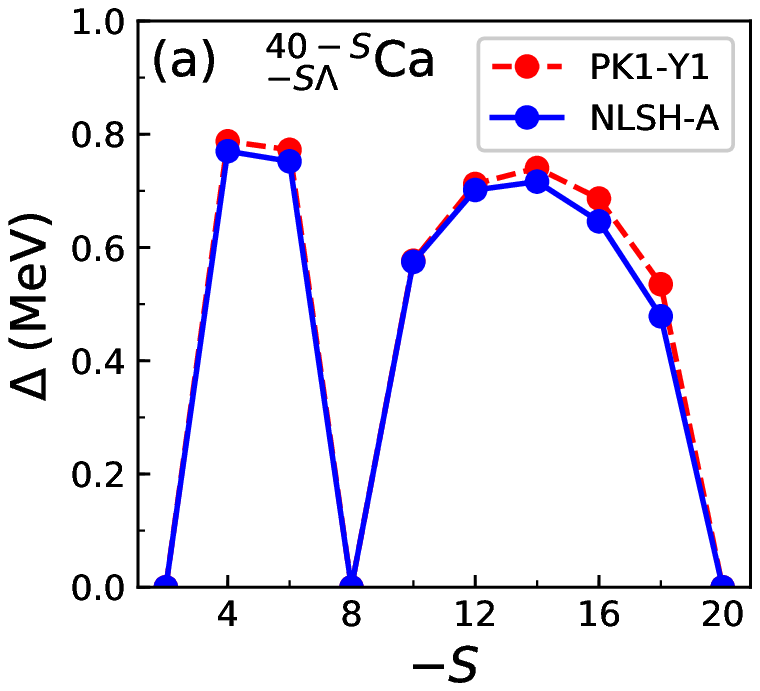}
\includegraphics[width=0.32\textwidth]{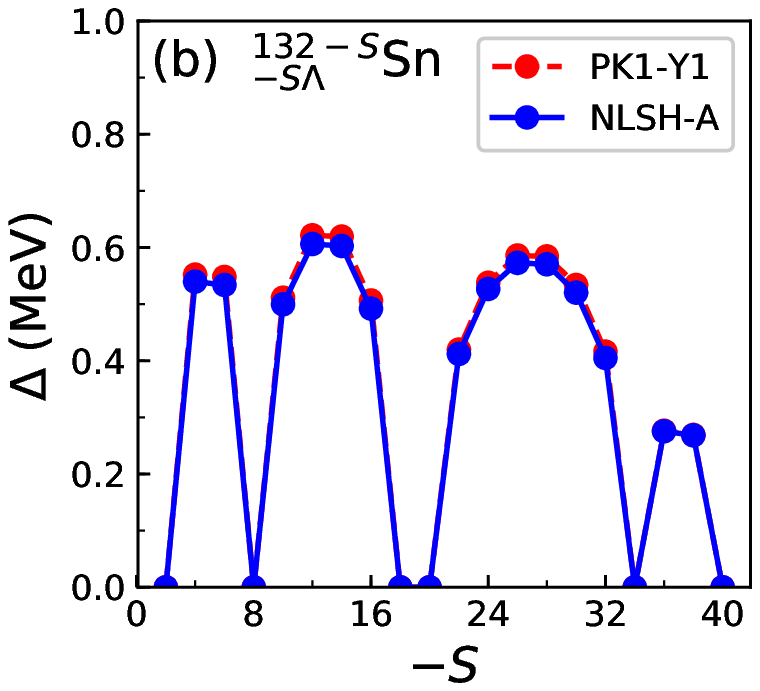}
\includegraphics[width=0.32\textwidth]{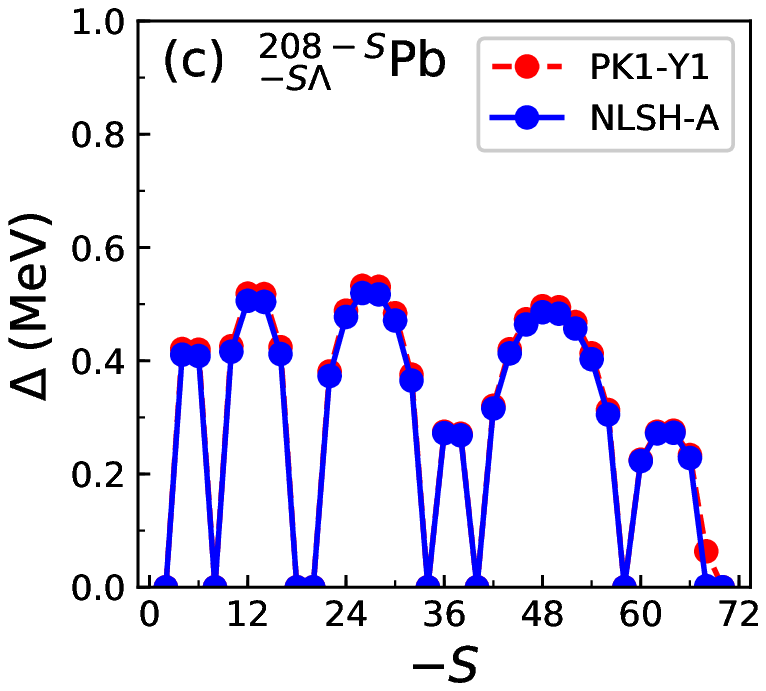}
\caption{(Color online) 
The pairing gap of $\Lambda$s as a function of the strangeness number $-S$ for
(a)  $^{40-S}_{-S\Lambda}$Ca ($-S=$ 0--20), 
(b) $^{132-S}_{-S\Lambda}$Sn ($-S=$ 0--40) and 
(c) $^{208-S}_{-S\Lambda}$Pb ($-S=$ 0--70) obtained in the MDC-RHB calculations. 
The effective interactions PK1-Y1 and NLSH-A are used for the $ph$ channel and 
the separable pairing force of finite range with the pairing strength 
$G_\Lambda= 4/9 \cdot G_N = 323.56$ MeV$\cdot$fm$^3$ is used for $pp$ channel.  
}\label{fig:gap_func_sn}
\end{figure}

The pairing gap is one of the typical quantities to characterize pairing effects.
We have calculated the average pairing gap as \cite{Bender2000_EPJA8-59}
\begin{equation}
 \Delta_\Lambda 
 = \sum_k \left\langle u_k v_k \Delta_k \right\rangle 
   \bigg/
   \sum_k \left\langle u_k v_k \right\rangle,
\end{equation}
where $\Delta_k$ is the pairing gap corresponding to a single $\Lambda$ state
$k$ in the canonical basis and $u^2_k$ and $v^2_k$ give the empty and occupation
probabilities, respectively.
Figure~\ref{fig:gap_func_sn} shows the $\Lambda\Lambda$ pairing gaps of 
 $^{40-S}_{-S\Lambda}$Ca, 
$^{132-S}_{-S\Lambda}$Sn and 
$^{208-S}_{-S\Lambda}$Pb. 
It can be seen that for almost every hypernucleus, the pairing gaps of $\Lambda$s 
obtained from PK1-Y1 and NLSH-A are very similar.
One can also find that the $\Lambda\Lambda$ pairing gaps are zero when the strangeness 
number is 2, 8, 20, 34, 40, 58 and 70, consistent with the conclusion drawn from
the two-Lambda separation energies that they are magic numbers for $\Lambda$s. 
There is a clear dependence of $\Delta_\Lambda$ on the mass number of the core nucleus.
For
$^{40-S}_{-S\Lambda}$Ca, 
$^{132-S}_{-S\Lambda}$Sn and 
$^{208-S}_{-S\Lambda}$Pb, the maximal values of $\Delta_\Lambda$ are
a bit smaller than 0.8 MeV, around 0.6 MeV and smaller than 0.6 MeV, respectively.
That is, the heavier the core, the smaller the pairing gap of $\Lambda$s.
This dependence is consistent with the observation 
in normal open shell nuclei that the pairing gap for nucleons decreases
with the mass number.
We will discuss more about this dependence later.
%can be described by the empirical formula $\Delta_{N} \approx 12 A^{-1/2}$ MeV \cite{Bohr1969_Nucl_Structure}.
% 46^{1/2} = 0.1474419561549
%160^{1/2} = 0.079056941504209
%272^{1/2} = 0.060633906259083
Meanwhile, when comparing the maximal values of $\Delta_\Lambda$ with
$\Delta_{N}$, one may also notice that the pairing 
effects of $\Lambda$s are weaker than nucleons; 
e.g., $\Delta_\Lambda <$ 0.8 MeV for $^{44}_{4\Lambda}$Ca, while for an open
shell nucleus with $A=44$, $\Delta_N$ is about 1.8 MeV according to the 
empirical formula $\Delta_N \approx 12 A^{-1/2}$ MeV \cite{Bohr1969_Nucl_Structure}.
Since the pairing strength for $\Lambda$s has been taken as 4/9 of that for 
nucleons, it is not unexpected that the pairing effects of $\Lambda$s are 
weaker compared to nucleons.

\begin{figure}[!htbp]
\centering
\includegraphics[width=0.48\textwidth]{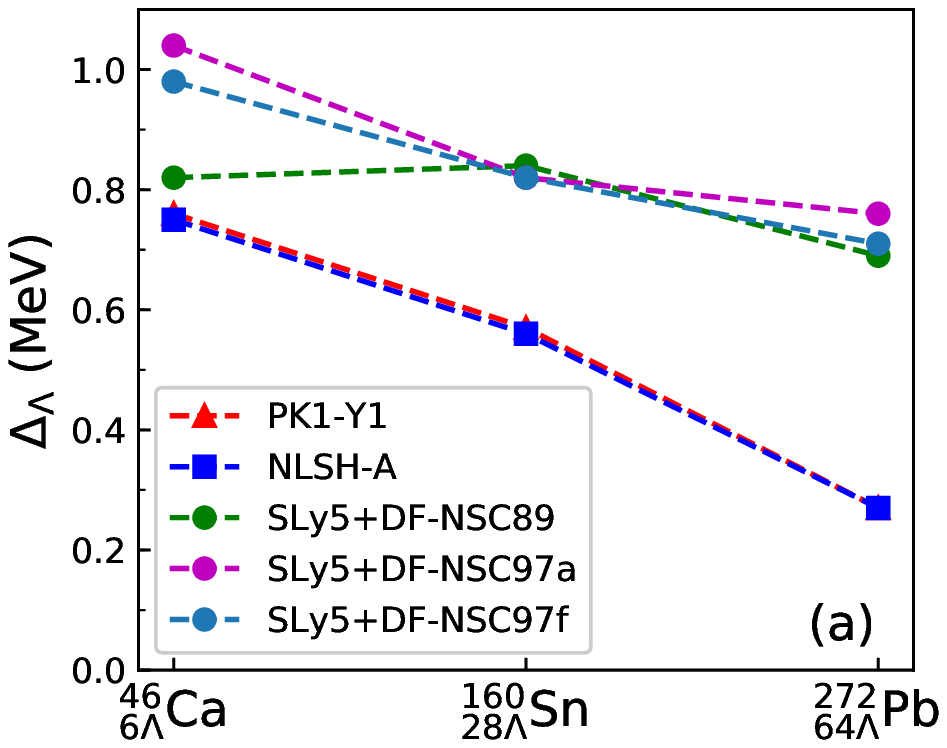}
\includegraphics[width=0.48\textwidth]{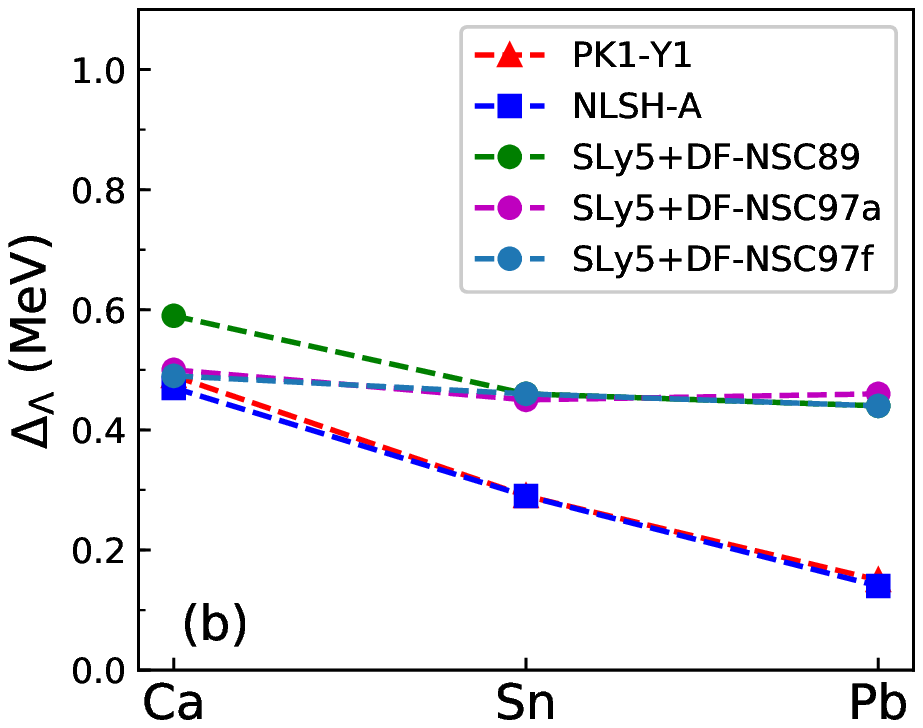}
\caption{(Color online) 
(a) $\Lambda\Lambda$ pairing gaps for 
 $^{46}_{6\Lambda}$Ca, 
$^{160}_{28\Lambda}$Sn and 
$^{272}_{64\Lambda}$Pb and 
(b) the average $\Lambda\Lambda$ pairing gaps as defined 
in Eq.~(\ref{eq:gap_ave}) for 
 $^{40-S}_{-S\Lambda}$Ca ($-S = $  6--20),
$^{132-S}_{-S\Lambda}$Sn ($-S = $ 18--40) and
$^{208-S}_{-S\Lambda}$Pb ($-S = $ 58--70) 
compared with the HFB results \cite{Gueven2018_PRC98-014318}
with SLy5 for the $NN$ interaction and DF-NSC89, DF-NSC97a and DF-NSC97f 
for the $\Lambda N$ interaction in the $ph$ channel. 
In the MDC-RHB calculations, the effective interactions PK1-Y1 and NLSH-A 
are used for the $ph$ channel and the separable pairing force of finite 
range with the pairing strength $G_\Lambda=4/9 \cdot G_N=323.56$ MeV$\cdot$fm$^3$
is used for $pp$ channel.
}\label{fig:Lamb_pairing_gap}
\end{figure}

There have not been much work on the pairing effects of hyperons in finite 
nuclei, though some efforts were made to the study of double-$\Lambda$ 
and multistrangeness ($-S\ge 3$) hypernuclei 
\cite{Miyahara1983_PTP69-1717,%Shen1998_APSOE7-258,
Lu2002_CPL19-1775,
Lu2008_CPL25-3613,Shoeb2009_JPG36-045104,
Gal2011_PLB701-342,
Margueron2017_PRC96-054317,
Gueven2018_PRC98-014318,
Tanimura2019_PRC99-034324}.
In Ref.~\cite{Gueven2018_PRC98-014318}, G\"uven et al. have investigated
multistrangeness hypernuclei with the Hartree-Fock-Bogoliubov (HFB) theory
and obtained interesting results concerning pairing effects of $\Lambda$s.
Next we make a brief comparison of our results with Ref.~\cite{Gueven2018_PRC98-014318}.
In Fig.~\ref{fig:Lamb_pairing_gap}(a), the pairing gaps for $\Lambda$s in 
three typical multistrangeness hypernuclei
 $^{46}_{6\Lambda}$Ca, 
$^{160}_{28\Lambda}$Sn and 
$^{272}_{64\Lambda}$Pb 
are compared with the HFB results \cite{Gueven2018_PRC98-014318}.
It can be seen that $\Lambda\Lambda$ pairing gaps of these three nuclei in 
the present work are smaller than those given in Ref.~\cite{Gueven2018_PRC98-014318}. 
Furthermore, the $\Delta_\Lambda$ from the MDC-RHB theory
decreases faster with $A$ than the HFB predictions. 
This conclusion holds also for the average pairing gap
\begin{equation}
 {\bar{\Delta}_\Lambda} \equiv 
 \frac{1}{m} \sum_{-S} \Delta_{\Lambda}(^{A-S}_{-S\Lambda}\mathrm{X}),
 \ \ \mathrm {X\ =\ Ca, Sn\ and\ Pb},  
 \label{eq:gap_ave}
\end{equation}
%where the summation is over $-S=6$--20 for Ca ($m=$ 8), 
%$-S=18$--40 for Sn ($m=$ 12) and $-S=58$--70 for Pb ($m=$ 12).
for 
 $^{40-S}_{-S\Lambda}$Ca ($-S=6$--20 and $m=$ 8), 
$^{132-S}_{-S\Lambda}$Sn ($-S=18$--40 and $m=$ 12) and 
$^{208-S}_{-S\Lambda}$Pb ($-S=58$--70 and $m=$ 12),
as seen in Fig.~\ref{fig:Lamb_pairing_gap}(b).
In Ref.~\cite{Gueven2018_PRC98-014318}, a zero-range $\delta$ force is adopted 
for the $\Lambda\Lambda$ pairing and its strength for 
$^{40-S}_{-S\Lambda}$Ca, $^{132-S}_{-S\Lambda}$Sn and 
$^{208-S}_{-S\Lambda}$Pb has been adjusted separately within a sharply 
truncated pairing window.
The adjustment was made by fitting the average pairing gap (\ref{eq:gap_ave}) 
to the maximal pairing gap in uniform hypernuclear matter given in 
Ref.~\cite{Tanigawa2003_PRC68-015801} 
at a certain density corresponding to the averaged density of
$^{40-S}_{-S\Lambda}$Ca ($-S=6$--20), 
$^{132-S}_{-S\Lambda}$Sn ($-S=18$--40) and 
$^{208-S}_{-S\Lambda}$Pb ($-S=58$--70), respectively. 
Thus for Ca, Sn and Pb isotopes, the pairing strengths are different and
the pairing interaction is always the strongest for Pb, as seen
in TABLE IV of Ref.~\cite{Gueven2018_PRC98-014318}. 
In the present work, however, a global finite-range pairing force 
[Eq.~(\ref{eq:separable})] is adopted and, thus, there is no hard cut-off for 
the pairing window. 

In normal nuclei, it has been well known that the pairing gap 
%(corresponding to the odd-even mass difference for well-deformed nuclei) 
for nucleons $\Delta_{N}$ declines more or less with 
$\sqrt{A}$ \cite{Bohr1969_Nucl_Structure}.
This decreasing tendency is roughly consistent with the dependence of 
$\Lambda\Lambda$ pairing gaps with respect to the number of $\Lambda$s 
obtained in the present work. 
Nevertheless, in Ref.~\cite{Gueven2018_PRC98-014318}, the decrease of 
the $\Lambda\Lambda$ pairing gaps is much gentler with the number of $\Lambda$s. 
This is quite interesting and should be investigated further.
 
To summarize, we have proposed a relationship between effective pairing 
interactions for hyperons and for nucleons based on the quark model.
Namely, the ratio between the strength of the effective pairing interaction 
for the hyperon $Y$ consisting of $n_{\mathrm{u/d}}$ u/d quarks and that for
the nucleon is $n^2_{\mathrm{u/d}}/9$. 
For $\Lambda$s, this ratio is simply $4/9$. 
A separable pairing force of finite range has been implemented in 
the MDC-RHB theory to investigate $\Lambda\Lambda$ pairing effects in 
multi-$\Lambda$ Ca, Sn and Pb hypernuclei.
By examining the two-$\Lambda$ separation energy $S_{2\Lambda}$ and 
the pairing gap $\Delta_\Lambda$, 
it is revealed that $-S =$ 2, 8, 20, 34, 40 and 58 are magic or semi-magic 
numbers for $\Lambda$s. 
The $\Delta_\Lambda$ decreases when the mass number of the core nucleus increasing.
It is also found that the pairing effects of $\Lambda$s are weaker than 
nucleons due to the supression of the pairing strength by a factor of 4/9. 

Finally, let us make two further remarks.
First, one may notice that the ratio $n^2_{\mathrm{u/d}}/9$ is probably very rough.
Other factors such as the violation of the OZI rule 
\cite{Kuwabara1995_PTP94-1163,Meissner1997_PLB408-381}, 
medium effects \cite{Tsushima1997_PLB411-9}, possible different couplings of 
$\rho$ to baryons \cite{Saito1996_NPA609-339} and mass splittings for baryons 
\cite{Noble1980_PLB89-325} may alternate this ratio or make the relation more 
complex between pairing interactions of $\Lambda$s and nucleons. 
Second, although in the present work the ratio $n^2_{\mathrm{u/d}}/9$ between 
the strength of the pairing interaction for hyperons and that for nucleons 
has been used in the framework of the RMF models, we expect it is also applicable 
to non-relativistic mean field models. 

\section*{Acnowledgements}

Helpful discussions with Xiang-Xiang Sun and Kun Wang are gratefully acknowledged. 
This work has been supported by 
the National Key R\&D Program of China (2018YFA0404402),
the NSF of China (11525524, 11621131001, 11947302, 11975031, and 11961141004),
the CAS Key Research Program of Frontier Sciences (QYZDB-SSWSYS013),
the CAS Key Research Program (XDPB09-02),
the Inter-Governmental S\&T Cooperation Project between China and Croatia,
and
the IAEA CRP ``F41033''. 
The computation of this work was supported by 
the HPC Cluster of KLTP/ITP-CAS and 
the Supercomputing Center, CNIC of CAS.

\section*{References}

\bibliographystyle{elsarticle-num_20200224-sgzhou}
%\bibliographystyle{apsrev4-1}
%\bibliography{../../Notes/bib/ref}
\bibliography{../../../information/refs/JabRef/sgzhou}

\end{document}